%% file: openiss-rs-kg-framework.tex
\documentclass[preprint,12pt]{elsarticle}

\usepackage{amssymb}

\usepackage[ruled,vlined,linesnumbered,resetcount]{algorithm2e}
\DontPrintSemicolon 
\newcommand{\AlgoResetCount}{\renewcommand{\@ResetCounterIfNeeded}{\setcounter{AlgoLine}{0}}}
\newcommand{\AlgoNoResetCount}{\renewcommand{\@ResetCounterIfNeeded}{}}
\newcounter{AlgoSavedLineCount}

\makeatother

\usepackage{url}
\usepackage{multirow}

\usepackage[english]{babel}
\usepackage{hyperref}
\addto\extrasenglish{%
}

\input{commands}

\journal{Science of Computer Programming}

\begin{document}

\begin{frontmatter}



\title{Application of Knowledge Graphs to Provide Side Information for Improved Recommendation Accuracy}


\author{Yuhao Mao, Serguei A. Mokhov, Sudhir Mudur}

\affiliation{organization={Computer Science and Software Engineerin, Concordia University},
            addressline={1515, Ste Catherine West}, 
            city={Montreal},
            postcode={H3G 2W1}, 
            state={Quebec},
            country={Canada}}

\begin{abstract}
Personalized recommendations are popular in these days of Internet driven activities, specifically shopping.
Recommendation methods can be grouped into three major categories, content based filtering,
collaborative filtering and machine learning enhanced. Information about products and preferences
of different users are primarily used to infer preferences for a specific user. Inadequate information
can obviously cause these methods to fail or perform poorly. The more information we provide to these methods,
the more likely it is that the methods perform better. Knowledge graphs represent the current trend in recording
information in the form of relations between entities, and can provide additional (side) information
about products and users. Such information can be used to improve nearest neighbour search,
clustering users and products, or train the neural network, when one is used. In this work,
we present a new generic recommendation systems framework, that integrates knowledge graphs
into the recommendation pipeline. We describe its software design and implementation,
and then show through experiments, how such a framework can be specialized for a domain,
say movie recommendations, and the improvements in recommendation results possible
due to side information obtained from knowledge graphs representation of such information.
Our framework supports different knowledge graph representation formats, and facilitates format conversion, merging and information extraction needed for training recommendation methods.    
\end{abstract}




\end{frontmatter}



\section{Introduction}
\label{sec:intro-essentialbg}

Recommendation systems are the product of the rapid development of the Internet. The use of the Internet to support different human activities has been on a constant rise, but propelled by the current pandemic, we start to see a very steep rise. There are more and more product offerings and providers on the Internet and user acceptance has jumped. However, the users are over loaded with information. In such situations, the recommendation system comes into play \cite{adomavicius05:recSysSurvey}. The recommendation system is essentially a technical means for users to narrow the information they are interested in from the massive amount of information available on the Internet, when the user desired product is not specific to a single item \cite{wikipedia-recommendation-system}. 

Applications of recommendation systems are very wide.
According to reports, the recommendation system has brought 35\% of sales revenue to Amazon \cite{increase-rs-amzon}
and up to 75\% of consumption to Netflix \cite{increase-rs-netflix}, 
and 60\% of the browsing on the Youtube homepage comes from recommendation services \cite{increase-rs-youtube}.
It is also widely used by various Internet companies. As long as the company has a large number of products to offer to the clients, the recommendation system will be useful \cite{heist2020knowledge,chah2018ok,hanika2019discovering}. The current application fields of the recommendation system have transcended beyond e-commerce, into news, video, music, dating, health, education, etc. 

A recommendation system can be regarded as an information filtering system, 
which can learn the user's interests and preferences based on the user's files or historical behaviour, 
and predict the user's rating or preference for a given item, based on information about the item, and the user. Clearly, the more information, the recommendation method has about users and products, the better is its ability to predict user preferences. 

In this paper, we present a new generic software framework which enables easy integration of any available additional information, called side information, by including it into a knowledge graph to be used for training the recommendation method.  

In most recommendation scenarios,
items may have rich associated knowledge in the form of interlinked information, 
and the network structure that depicts this knowledge is called a knowledge graph. 
Information is encoded in a data structure called triples made of subject-predicate-object statements. A knowledge graph on the item side greatly increases information about the item, 
strengthens the connection between items, provides a rich reference value for the recommendation, 
and can bring additional diversity and interpretability to the recommendation result. 
recommendation systems.

There is a clear need for a general framework, which (i) integrates search and update of information, (ii) includes crawling of websites for additional information, (iii) supports storing of the information in a structured, easily accessible manner, (iv) enables easy retrieval of the information about items and users as input for the training of a recommendation system. Adding a knowledge graph into the recommendation framework can help us better manage knowledge data, process data, and query the information we need faster.

\textbf{Firstly}, knowledge graphs as a form of structured human knowledge have drawn great research attention from both the academia and the industry \cite{dong2014knowledge,nickel2015review,wang2017knowledge}.
A knowledge graph is a structured representation of facts, consisting of entities, relationships, 
and semantic descriptions.
The knowledge graph can be used wherever there is a relationship. 
It has successfully captured a large number of customers,
including Walmart, Google, LinkedIn, Adidas, HP, FT Financial Times, etc. 
well\-known companies and institutions.  Applications are still growing. 

Compared with traditional data bases and information retrieval methods, the advantages of knowledge graph are the following:

\begin{itemize}
    \item {Strong ability to express relationships}: Based on graph theory and probability graph models, it can handle complex and diverse association analyses.
    \item {Knowledge learning}: it can support learning functions based on interactive actions such as reasoning, error correction, and annotation, and continuously accumulates knowledge logic and models, improves system intelligence.
    \item {High-speed feedback}: Schematic data storage method enables fast data retrieval speeds. 

\end{itemize}

Knowledge graphs usually have two main types of storage formats \cite{zhao2018architecture,auer2018towards}: one is RDF (Resource Description Framework) based storage, and the other is graph database (e.e., neo4j), with their advantages and disadvantages.

\textbf{Secondly}, the recommendation system 
has become a relatively independent research direction. 
It is generally considered to have started with the 
GroupLens system launched by the GroupLens research group of the University of Minnesota in 1994 \cite{ekstrand2011collaborative}. 
As a highly readable external knowledge carrier, knowledge graphs provide a great possibility to improve algorithm interpretation capabilities \cite{guo2020survey}. Therefore, combining knowledge graph with recommendation systems is one of the hottest topics in the current recommendation system research.

Our main contribution is the following:

\begin{itemize}
    \item We present an overall architecture that allows users to build knowledge graphs, 
    display knowledge graphs, and enable recommender algorithms to be trained with knowledge when predicting user ratings for items. We demonstrate our framework for the movie recommendation domain.
 \end{itemize}
 Other contributions include:
 \begin{itemize}
    
    
    \item A pipeline architecture that allows users to crawl data, 
    build and merge knowledge graphs in different formats, to display knowledge graphs and extract information, without knowing the underlying format details.
    
    \item We offer a way for the recommendation systems researchers to enable 
    recommendation system experiments on top of \tool{TensorFlow} and \tool{Keras}.
\end{itemize}

The rest of the paper is organized as follows. In the next section we provide a brief review of existing software frameworks for recommendations reported in the literature. Next, we describe the design and implementation of our  framework in a top-down manner, along with examples of instantation and applications. Lastly, we present our conclusions and some extensions.

\section{Literature Review}
\label{sec:rsframeworks}

While we can find a tremendous lot on recommendation algorithms, a literature scan reveals only a few attempts at development of frameworks for recommendation systems. Below is a brief review of these. 

\subsection{A Gradient based Adaptive Learning Framework for Efficient Personal Recommendation}
Yue et al. \cite{ning2017gradient} use gradient descent to learn the user's model for the recommendation. 
Three machine learning algorithms (including logistic regression, 
gradient boosting decision tree and matrix decomposition) are used. 
Although gradient boosting decision tree can prevent overfitting and has strong interpretability, 
it is not suitable for high-dimensional sparse features, usually the case with items and users. 
If there are many features, each regression tree will consume a lot of time.

\subsection{Raccoon Recommendation Engine}
\label{sec:recoon}

Raccoon \cite{recommendation-Raccoon} is a recommendation system framework based on collaborative filtering. 
The system uses k-nearest-neighbours to classify data. 
Raccoon needs to calculate the similarity of users or items.  The original implementation of Raccoon uses Pearson which was good for measuring similarity of discrete values in a small range. But to make the calculation faster, one can also use Jaccard,  which is a calculation method for measuring binary rating data (ie like/dislike).
But the collaborative filtering algorithm does not care about the inner connection of characters or objects. 
It only uses the user ID and product ID to make recommendations. 

\subsection{Good Enough Recommendations (GER)}
\label{sec:ger}

GER (Good Enough Recommendation) \cite{recommendation-ger} is a scalable, 
easy-to-use and easy-to-integrate recommendation engine. 
GER is an open source NPM module. 
Its core is the same as the knowledge graph triplet (people, actions, things).
GER recommends in two ways. 
One is by comparing two people, looking at their history, 
and another one is from a person's history.
Its core logic is implemented in an abstraction called the Event Storage Manager (ESM).
Data can be stored in memory ESM or PostgreSQL ESM. 
It also provides corresponding interfaces to the framework developer, 
including the Initialization API for operating namespace, 
the Events API for operating on triples, 
the Thing Recommendations API for computing things, 
the Person Recommendations API for recommending users, 
and Compacting API for compressing items.

\subsection{LensKit}

LensKit \cite{framework-lkpy} is an open-source recommendation system based on java, 
produced by the GroupLens Research team of the University of Minnesota. 
But the java version of LensKit has been deprecated, and the latest version uses python.
The python version of Lenskit is a set of tools for experimenting and researching recommendation systems.
It provides support for training, running and evaluating recommendation systems.
The recommendation algorithms in \api{LensKit} include SVD, 
Hierarchical Poisson Factorization, and KNN.
\api{LensKit} can work with any data in \api{pandas.DataFrame} with the expected (fixed set) columns. 
\api{Lenskit} loads data through the \api{dataLoader} function. 
Each data set class or function takes a path parameter specifying the location of the data set. 
These data files have normalized column names to fit with LensKit's general conventions.

\subsection{Deep Knowledge-Aware Network for News Recommendation (DKN)}
\label{sec:DKN}

DKN \cite{wang2018dkn} proposes a model that integrates the embedded representation of knowledge graph entities 
with neural networks for news recommendation. 
News is characterized by highly condensed representation and contains many knowledge entities, snd it is time sensitive. A good news recommendation algorithm 
should be able to make corresponding changes as users’ interests change. 
To solve the above problems, the DKN model is proposed. 
First, a knowledge-aware convolutional neural network (KCNN) is used to integrate the semantic representation of news with the knowledge representation to form a new embedding, 
and then the attention from the user’s news click history to the candidate news is established. 
The news with higher scores is recommended to users.

\subsection{Multi-task Feature Learning for Knowledge Graph enhanced Recommendation (MKR)}
\label{sec:MKR}

MKR \cite{wang2019multitask} is a model that uses knowledge graph embedding tasks to assist recommendation tasks. 
These two tasks are linked by a cross-compression unit, 
which automatically shares potential features,
and learns the high-level interaction between the recommendation system 
and entities in the knowledge graph.
It is shown that the cross-compression unit has sufficient polynomial approximation ability, 
and MKR is a model of many typical recommendation systems 
and multi-task learning methods.

\subsection{Summary of Recommendation System Literature Review}
There are three types of recommendation system models 
(collaborative filtering, content-based and machine learning) in the above frameworks. 
\xt{tab:frameworkall} is a summary characterizing the above frameworks.

\begin{table}
	\resizebox{\textwidth}{15mm}{
	\begin{tabular}{c|c|c|c|c|c|c|c} 
		\textbf{Name} & \textbf{Type} & \textbf{Storage types} & \textbf{Domain} & \textbf{Example uses} & \textbf{Language} & \textbf{ML Models} & \textbf{Maintained}\\ \hline
		Yue Ning et. al. & Decision Tree & - &  content recommendation  & content recommendation  & -  & Boosting & -\\
 
        Racoon  & Collaborative Filtering & Redis  & cross-domain & ${https://github.com/guymorita/benchmark_raccoon_movielens}$  & Java  & KNN & Jan 10, 2017 \\
 
        GER  & Collaborative Filtering & PostgreSQL  & movie  & ${https://github.com/grahamjenson/ger/tree/master/examples}$  & Java  & - & Jul 9, 2015 \\
        
        LensKit  & Machine Learning & LocalFile  & cross-domain  & ${https://github.com/lenskit/lkpy/tree/master/examples}$  & Python  & SVD, hpf & Nov 10, 2020 \\
        
        DKN  & knowledge graph enhanced  & LocalFile  & News  & ${https://github.com/hwwang55/DKN}$  & Python  & tensorflow & Nov 22, 2019 \\
        
        MKR  & knowledge graph enhanced  & LocalFile  & cross-domain  & ${https://github.com/hwwang55/MKR}$  & Python  & tensorflow & Nov 22, 2019 \\
        
        PredictionIO  & machine learning  & Hadoop,HBase  & -  & ${https://github.com/apache/predictionio}$  & Scala  & Apache Spark MLlib & Mar 11, 2019 \\
        
        Surprise & Collaborative Filtering & LocalFile  & movie,joke  & ${https://github.com/NicolasHug/Surprise/tree/master/examples}$  & Python  & matrix factorization, KNN & Aug 6, 2020 \\
		\hline
	\end{tabular}}
	\caption{Characteristics of Existing Frameworks}
	\label{tab:frameworkall}
\end{table}

As we can see there are no recommendation system software frameworks yet that are generic to support web crawling, information update, visualization and input to recommendation methods independent of storage formats and algorithms. We will briefly discuss the limitations of presently available frameworks. Yue et al. mainly uses the boosting model, which makes the training time high. Because Raccoon uses the K-nearest-neighbours model, it cannot handle new users or new items, well known as the cold start problem. Further, it also has data sparseness and scalability issues. The advantage of GER is that it contains no business rules, with limited configuration, and almost no setup required. But this is at the expense of the scalability of the engine. Other limitations are that it does not generate recommendations for a person with less history, and has the data set compression limit problem, i.e., certain items will never be used.  For example, if items are older or belong to users with a long history, these items will not be used in any calculations but will occupy space. The advantage of LensKit is that its framework contains many algorithms, such as funksvd and KNN, and users can call different algorithms according to their needs.  In LensKit the data is called through the path parameter, and the data has a specific format. This means that the LensKit framework can only process user, item and rating data. If the user has new data, such as user information or item classification, the LensKit framework cannot handle it. Although DKN uses knowledge graph embedding as an aid, it can only be used for news recommendations. The main disadvantage of MKR is that it is not a generic framework, and it inputs text documents as knowledge graphs, but not in their graph structured form, making it cumbersome to update knowledge.

\section{Framework Design}

\xf{fig:framework_overall} 
shows the overall design of our framework.  It is designed as a pipeline of tasks from end user input to final recommendations to the end user. Further, each stage in the pipeline is designed as a sub-framework, some stages are nested, enabling specialization and expansion at a more granular level. We denote a generic component as a frozen spot and its specialized component as a hot spot.
\begin{figure}[htbp]
    \begin{center}
    \includegraphics[scale=0.5]{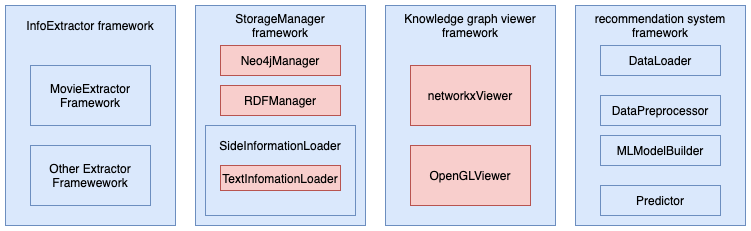}
    \end{center}
    \caption{Core components of the framework, each box in blue means the framework's frozen spot 
    and each box in red represents a set of the hot spots for each specialized framework.}
    \label{fig:framework_overall}
\end{figure}

The four major stages in the pipeline have the following functionality:
\begin{itemize}
    \item InfoExtractor framework: abstracts a web extractor.
    It takes a URL address (such as \file{.html}, \file{.htm}, \file{.asp}, \file{.aspx}, \file{.php}, \file{.jsp}, \file{.jspx}) 
    and extraction rules as input then formats the output that is extracted. 
    This is further nested, to enable domain level (movies, news, etc.) specialization.    
    
    \item StorageManager framework: abstracts different knowledge graph storage formats.
    It takes string data stream as input then generates the output file according to the required format.
     
    \item Knowledge graph viewer framework: abstracts knowledge graph visualization. 
    It takes triples stream as input then creates the visualization.
      
    \item Recommendation method: abstracts the recommendation method and knowledge input. The recommendation method can take these knowledge graph triples as input for training the recommendation prediction model.
\end{itemize}

\subsection{InfoExtractor Framework Design}
\label{sec:info-extractor-design}
It is necessary to abstract the common methods of information extractors. 
We will take the example of extracting movie information. 
We have designed a small module called \api{MovieExtractor} nested in \api{InfoExtractor}. 
\api{MovieExtractor} serves as a frozen spot to provide users with functions such as capturing movie information, as shown in \xf{fig:IMDB_extractor_framework}. 
\begin{figure}
    \begin{center}
    \includegraphics[scale=0.5]{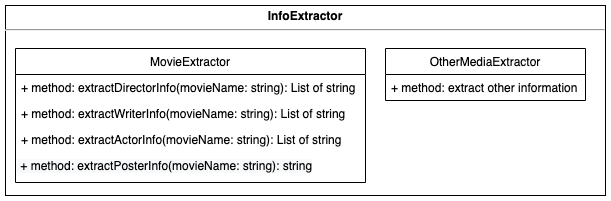}
    \end{center}
    \caption{Design of the InfoExtractor Framework.}
    \label{fig:IMDB_extractor_framework}
\end{figure}

\api{MovieExtractor} is the abstract class of the extractor, it has three basic methods, 
\api{extractDirectorInfo},\api{extractWriterInfo} and \api{extractActorInfo}. 
It is used to extract director information, author information and star information respectively.
\api{extractPosterInfo} is the abstract class of the poster downloader. 
It downloads the movie poster (an example of side information) and 
then converts the image into a string. to facilitate storage and coding.

There are many websites on the Internet that store information about movies, 
such as IMDB, Wikipedia, Netflix and Douban. 
Extractors dedicated to these websites can be used as hotspots to access our \api{MovieExtractor} API.

\subsection{StorageManager Framework Design}
\label{sec:kg-storage-design}

\api{StorageManager} Framework serves as a frozen spot, as shown in \xf{fig:kg_storage_fwdesign}. 
Also, the design allows us to add support for more kinds of storage modes easily without changing the frozen spots themselves.

\begin{figure}
    \begin{center}
    \includegraphics[scale=0.5]{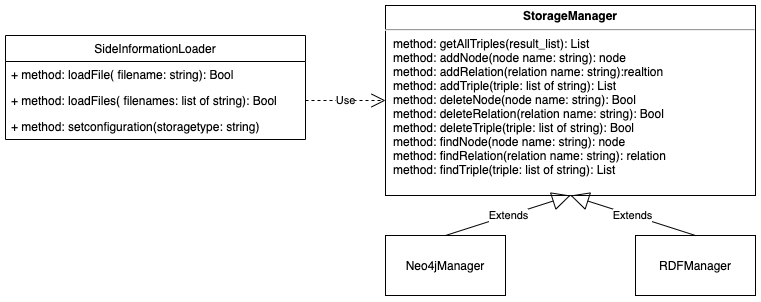}
    \end{center}
    \caption{Design of the knowledge graph StorageManager framework.}
    \label{fig:kg_storage_fwdesign}
\end{figure}

For the \api{SideInformationLoader},
We again design it as a frozen spot, 
and its responsibility is to add triples to the knowledge graph 
through the method in \api{StorageManager} to increase the richness of the knowledge provided to the recommender method.
\api{SideInfomationLoader} contains three methods, 
they are \api{loadFile}, \api{loadFiles} and \api{setConfiguration}.

\subsection{Knowledge Graph Viewer Framework Design}
\label{sec:kg-viewer-design} This frozen spot's, responsibility is to display triples for knowledge visualization. 
As shown in \xf{fig:kg_viewer_fwdesign}, \api{KnowledgeGraphViewer} is an abstract class, 
which contains a show method.
\begin{figure}
    \begin{center}
    \includegraphics[scale=0.6]{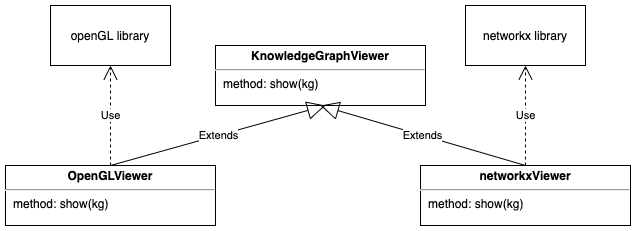}
    \end{center}
    \caption{Design of the knowledge graph viewer framework.}
    \label{fig:kg_viewer_fwdesign}
\end{figure}

\subsection{Recommendation Method Framework Design}
\label{sec:RS-framework-design}

We designed a dedicated framework for the recommendation method as shown in \xf{fig:rs_fwdesign}. 
\begin{figure}
    \begin{center}
    \includegraphics[scale=0.3]{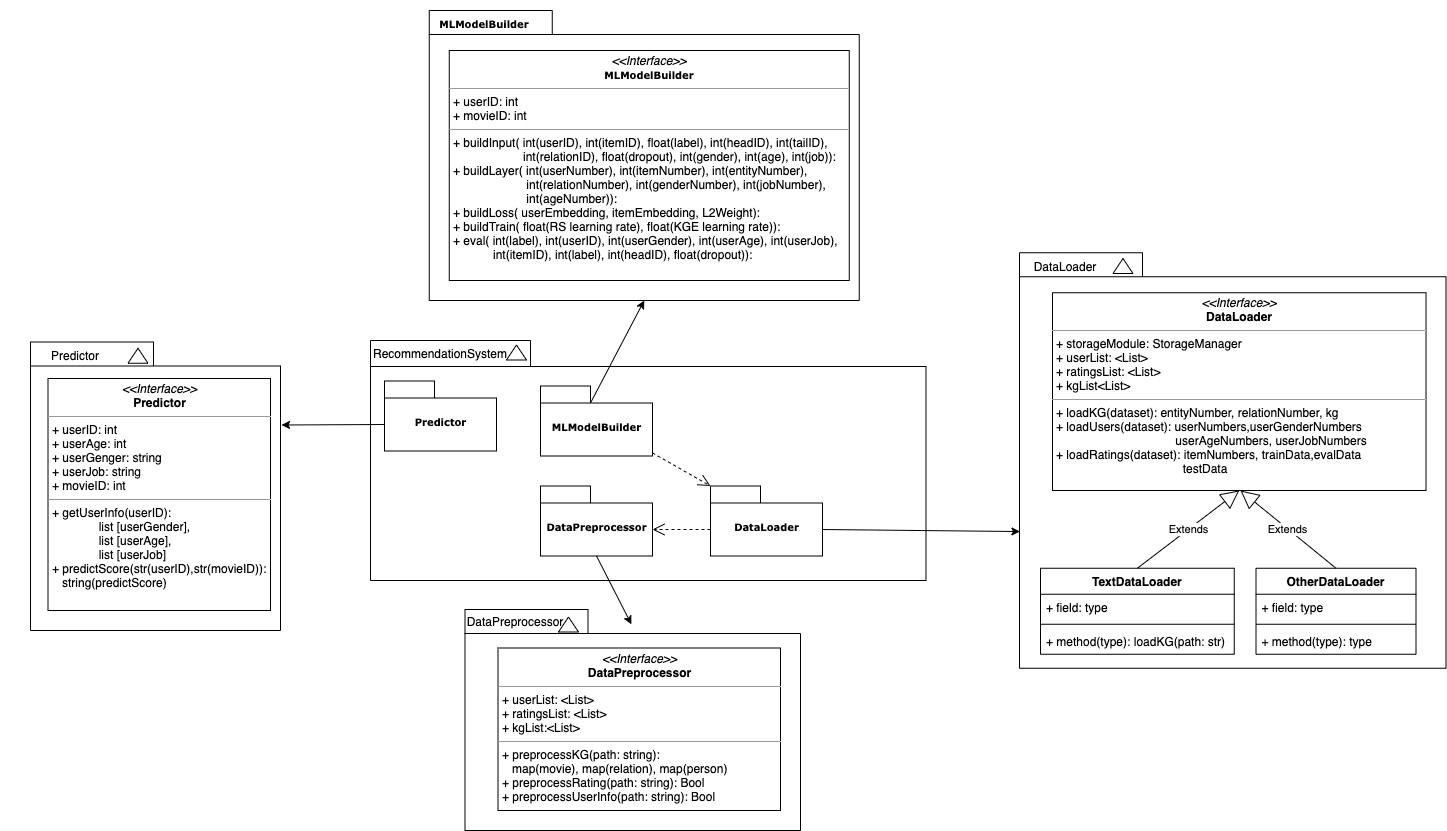}
    \end{center}
    \caption{Design of the recommendation method framework.}
    \label{fig:rs_fwdesign}
\end{figure}
It consists of four frozen spots.
Including \api{DataLoader} module, \api{DataPreprocessor} module, 
\api{MLModelBuilder} module and \api{Predictor} module.

\api{DataPreprocessor} module includes three methods, 
which are \api{preprocessKG}, \api{preprocessUserInfo} 
and \api{preprocessRating}.
\api{preprocessKG} is used to process knowledge graph triples. 
It returns three dictionaries to store the id corresponding to the product,
the id corresponding to the relationship 
and the id corresponding to the character.
\api{preprocessUserInfo} is used to encode user information, 
including the user's gender, age, and occupation. 
\api{preprocessRating} uses the three dictionaries stored in the previous step to convert the product and user ID in the rating information. 

The function of the \api{DataLoader} module is to read all the triples 
from the file generated in the previous step.
And return three lists, used to store the number of users, 
the number of items, the number of relations.
\api{loadKG} is used to load the KG file processed in the previous step,
calculate the number of entities, the number of relations, and return these values. 
These parameters are used to create the matrix of entities and relations when building the prediction model, for example neural network, in the next step.
\api{loadUsers} loads the user information file processed in the previous step, 
calculate the number of users, genders, ages and jobs. 
These parameters are used to create a user information matrix when building the neural network in the next step.
\api{loadRatings} is used to load the rating file, 
then calculate the number of items, 
and then divide the data into the training data, 
eval data and test data according to the ratio of 6:2:2. 
The model is built through \api{MLModelBuilder}.

After the user trains the model, s/he will get a pre-trained model. 
This model will be used to predict the user’s score in the later stage. 
\api{Predictor} framework is to facilitate this prediction. 
It includes \api{getUserInfo} and \api{predictScore} two methods.
If it is an old user, the user only needs to provide the user ID to query the user’s personal information. 
Three lists are returned, user’s gender, age, and job information.
\api{predictScore} returns a float value, 
which represents the user’s rating of the product.

\section{Framework Instantiation}

\subsection{IMDBExtractor Instantiation}
\label{sec:web-crawler-inst}

In \xs{sec:info-extractor-design}, we described the design of the \api{InfoExtractor} module within the framework. 
If instantiated for movie recommendation, we extract director, writer, stars information, movie genre, movie poster and other available movie data as side information. 

In \api{IMDBExtractor} module, we create a list for each kind of information to be extracted. 
Because there may be many directors, actors, and stars of the movie, 
the information for each category is returned as a list. 
If the relevant information cannot be found in IMDB, an empty list will be returned. 
Each triplet will be stored in the form of head, relation, tail. 

\subsection{StorageManager Instantiation}
\label{sec:kg-storage-inst}

In \xs{sec:kg-storage-design}, 
we proposed the design of frozen spot for a knowledge graph StorageManager framework in general. We created two storage modules as hot spots for knowledge graphs. 
They are \api{Neo4jManager} and \api{RDFManager}.
\xf{fig:instantiation_storage} shows the structure of this module.

\begin{figure}
    \centering
    \includegraphics[scale=0.4]{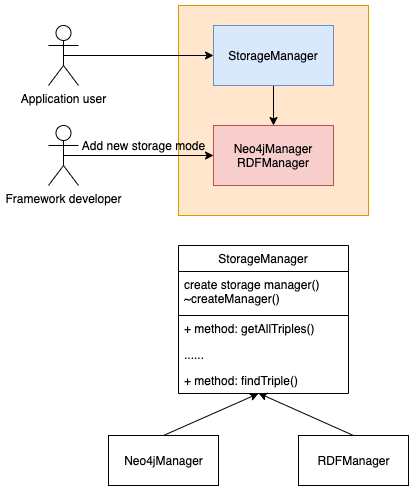}
    \caption{Instantiation of the StorageManager in framework.}
    \label{fig:instantiation_storage}
\end{figure}

\subsubsection{Neo4jManager}

To facilitate users to use the Neo4jManager framework, 
we provide a framework API for Neo4j operations. 
\xf{fig:structure_neo4j} illustrates the structure of Neo4j storage.
\begin{figure}
    \centering
    \includegraphics[scale=0.5]{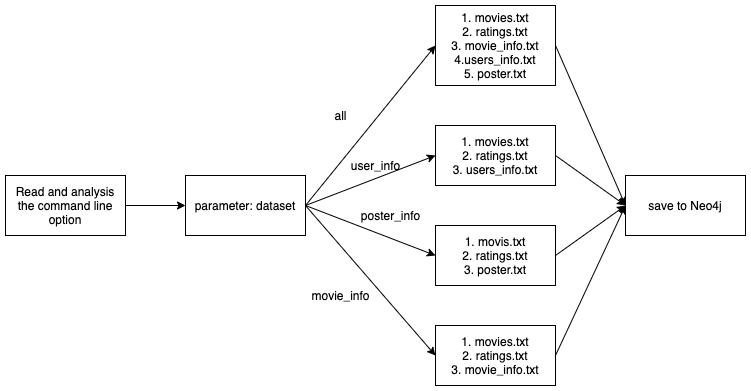}
    \caption{Structure of the Neo4j storage module in framework.}
    \label{fig:structure_neo4j}
\end{figure}

\subsubsection{RDFManager}
To facilitate users to use the RDFManager framework, 
we provide our framework’s API for all operations of owl, including:
\api{RDFSave}, \api{RDFGetOntology}, \api{RDFAddClass}, \api{RDFAddIndividual},
\api{RDFAddDataproperty}, 
\api{RDFAddDatapropertyValue} 
and \api{RDFAddObjectproperty}.
\xf{fig:structure_rdf} illustrates the structure of RDF storage.
When the user chooses to use RDF storage, it will call the RDFManager in the framework, 
use the API in our framework to operate on the triples, and then save it as an RDF file.
\begin{figure}
    \centering
    \includegraphics[scale=0.3]{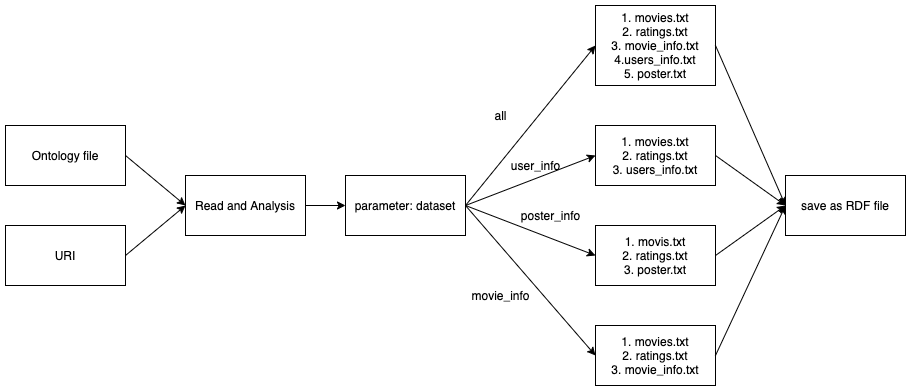}
    \caption{Structure of the RDF storage module in framework.}
    \label{fig:structure_rdf}
\end{figure}

\subsubsection{TextInfomationLoader}
In \xs{sec:kg-storage-design}, we described the design of the \api{SideInformationLoader}. 
To add side information to meet our requirements, 
based on the frozen spot of \api{SideInformationLoader}, 
we created a \api{TextInformationLoader} hot spot based on text documents. 
The overall idea is shown in \xf{fig:structure_textinfoloader}.
\begin{figure}
    \centering
    \includegraphics[scale=0.3]{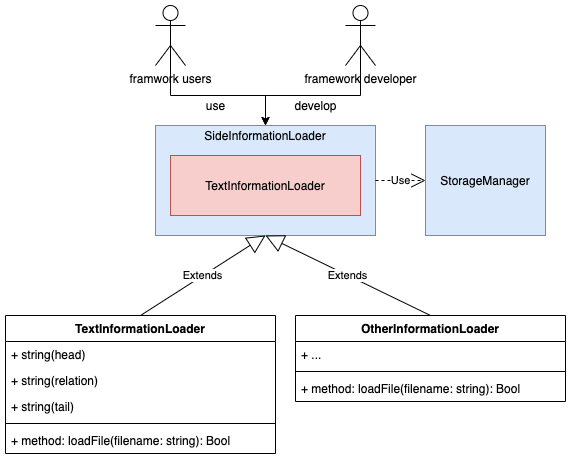}
    \caption{TextInformationLoader instantiation.}
    \label{fig:structure_textinfoloader}
\end{figure}
In our implementation, we need to read the file in the parameter, 
parse the file according to the format, and extract the head, relation and tail of the triples. 
Then add new triples to the knowledge graph through \api{addTriple} in \api{StorageManager}.
Since there is already a method for adding a single file, 
we only need to make some adjustments about \api{loadFiles}. 
When reading files, we just need to call \api{loadFile} for each file.

\subsection{networkxViewer Instantiation}
\label{sec:kg-viewer-inst}

In \xs{sec:kg-viewer-design}, we described the design of the viewer module. 
The workflow of the viewer module is shown in  \xf{fig:KG_viewer_workflow}.
\begin{figure}[htpb]
    \centering
    \includegraphics[scale=0.4]{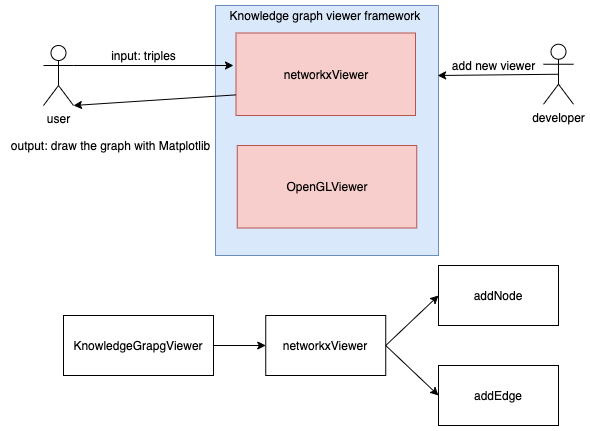}
    \caption{Instantiation of the networkxViewer framework.}
    \label{fig:KG_viewer_workflow}
\end{figure}
Our process can be described by the following steps:
\begin{enumerate}
    \item Read all the individual names and store all the individuals in \api{viewerIndividuals}.

    \item Take the individual list from the previous step and get all the information connected to this individual.
    
    \item According to the content in the individual, create nodes or links respectively. 
\end{enumerate}

\subsection{Recommendation Method Instantiation}
\label{sec:fw-inst-rs-framework}
The recommendation method is the most important part of our framework. 
In \xs{sec:RS-framework-design}, we proposed the design of a
generic recommendation method. 
Here we present an instance which adopts the deep learning method. 
We decided to implement the entire method ourselves, 
as we wanted to incorporate the benefits of side information obtained from using knowledge graphs. It is based on the work of \cite{45530,berg2017graph}. 
Our recommendation method is written in Python and built on top of TensorFlow.
\xf{fig:RS_framework_workflow} illustrates the workflow of the recommendation method module. 
\begin{figure}
    \centering
    \includegraphics[scale=0.3]{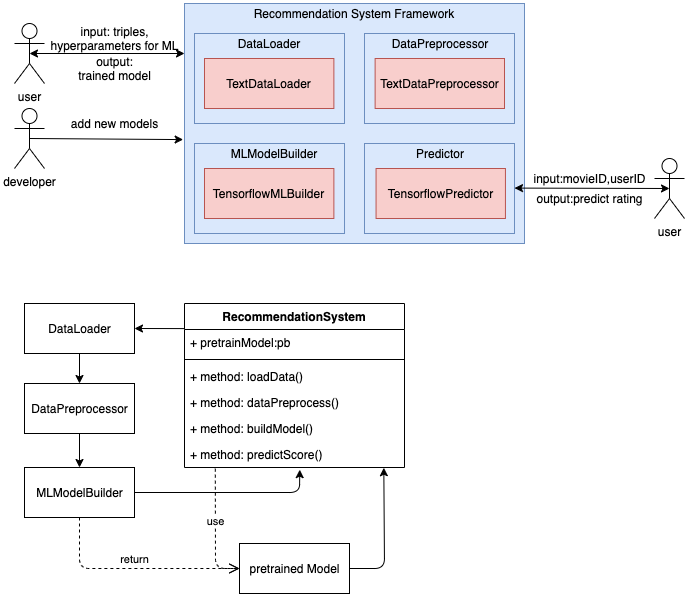}
    \caption{Workflow of recommendation system framework}
    \label{fig:RS_framework_workflow}
\end{figure}

We implemented and modified the network architecture shown in \xs{sec:MKR}.
For the recommendation method model, the structure is shown below in \xf{fig:structure_rs}.
In our implementation, we made some changes to the structure of the model.  
The original work does not contain user\_gender, user\_age and user\_job embeddings. 
Here we decided to add this as side information to improve the accuracy. 
To unify the latitude (in the deep learning network), we also choose arg.dim as the dimension of user\_age, user\_job and user\_gender.
\begin{figure}
    \centering
    \includegraphics[scale=0.2]{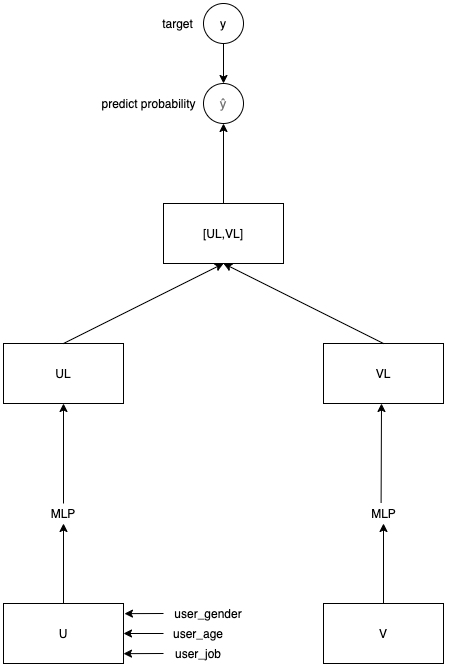}
    \caption{Structure of the recommendation module in RS framework.}
    \label{fig:structure_rs}
\end{figure}

For the knowledge graph embedding model, the structure is shown below in \xf{fig:structure_kge}.
Let's take our data as an example. There are $3883$ heads and four relations in the data.
Therefore, the item matrix is the $3883 \times \mathtt{arg.dim}$ matrix 
and the relation matrix is the $4 \times \mathtt{arg.dim}$ matrix.
Each time we take out the vector corresponding to the head 
from the head embedding according to the head index, 
and then process the crossover and compression unit to obtain an $H_l$b (head) vector.
The $R_l$ (relation) vector is obtained by looking up the vector of the relation index 
in the relation matrix and then passing through a fully connected layer.
Then we merge the [batch\_size, dim] dimension $H_l$ and $R_l$ into a [batch\_size, $2 \times \mathtt{arg.dim}$] vector, 
and this vector is passed through a fully connected layer to get a [batch\_size,arg.dim] vector.
This vector includes the predicted tail.
\begin{figure}
    \centering
    \includegraphics[scale=0.2]{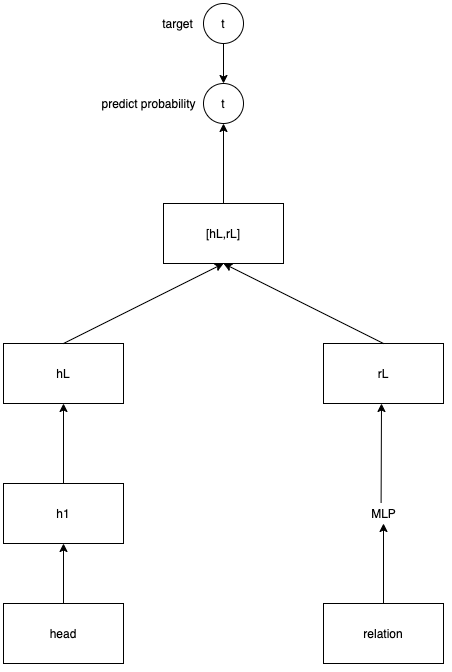}
    \caption{Structure of the knowledge graph embedding module in recommendation system framework.}
    \label{fig:structure_kge}
\end{figure}

For the cross and compress unit,
item and head are each a vector of dimension [batch\_size,arg.dim]. 
To facilitate calculation, we first expand them by one dimension 
so that they become [batch\_size,arg.dim,1] and [batch\_size,1,arg.dim] respectively, 
and then multiply them, to get the cross matrix c\_matrix, a matrix of [batch\_size,dim,dim], 
and a transpose matrix c\_matrix\_transpose, 
the two matrices are multiplied by different weights, 
then reshaped to obtain the final vectors.

During the training, our model is guided by three loss functions: 
$L_{RS}$ loss and $L_{KGE}$ loss and $L_{REG}$ loss.
	\begin{equation}
		L =  L_{RS} + L_{KG} + L_{REG}
	\end{equation}
The complete loss function is as follows:
\begin{itemize}
    \item {The first item is the loss in the recommendation module.}

    \item {The second item is the loss of the KGE module, 
    which aims to increase the score of the correct triplet and reduce the score of the wrong triplet.}
    
    \item {The third item is L2 regularization to prevent overfitting.}
\end{itemize}

\subsubsection{Predictor}
Once the model is trained, we need to make predictions. 
We use our predictor module.
There are two methods in this module, \api{getUserInfo} and \api{predictScore}.
\api{getUserInfo} is a method used to extract user's information. 

The model trained in the previous step is loaded into \api{predictScore}, 
and then different matrices are read by name. 
If the id entered by the user is greater than the dimension of the model. 
That means the id is a new user. 
Our recommendation will focus on the user's age and job.
Finally, a predicted float value is returned.

The steps we describe here can be represented by Algorithm \ref{algo:fw-app-prediction}.
 
\begin{algorithm}
    parser $\leftarrow$ init a argument parser\;
    dataset $\leftarrow$ init default dataset\;
    userid $\leftarrow$ init default  userid\;
    movieid $\leftarrow$ init default movieid\;
    \;
    \If {trained model path exist}
    {{load trained model}\;
    {predict user's rating to movie}\;}
    \Else
    {Error}
    \caption{prediction application procedure}
    \label{algo:fw-app-prediction}
\end{algorithm}

\subsection{Deployment}
\xf{fig:deployment_diagram} shows the diagram of all the required libraries.
In \api{InfoExtractor}, we need \api{request}, \api{bs4}, \api{IMDB} and \api{base64} libraries. 
The \api{requests} library is used to issue standard HTTP requests in Python. 
It abstracts the complexity behind the request into an API 
so that users can focus on interacting with the service and using data in the application.
HTML is composed of a ``tag tree", 
and the \api{bs4} library is a functional library responsible for parsing, traversing, and maintaining the ``tag tree".
\api{IMDB} is an online database of movie information.
\api{Base64} is a library that uses 64 characters to represent arbitrary binary data.

In \api{StorageManager}, we need \api{py2neo}, \api{owlready2} and \api{rdflib} libraries. 
\api{py2neo} can use Neo4j from within the Python application and from the command line.
\api{owlready2} is a module for ontology-oriented programming in Python.
\api{rdflib} is used to parse and serialize files in RDF, owl, JSON and other formats.

In \api{KGViewer}, we need \api{networkx} and \api{matplotlib} libraries. 
\api{networkx} is a graph theory and complex network modelling tool developed in Python language, 
which can facilitate complex network data analysis and simulation modelling.
The \api{matplotlib} library is an essential data visualization tool.

In \api{RecommendationSystem}, we need \api{random}, \api{numpy}, \api{sklearn}
\api{linecache} and \api{tensorflow} libraries. 
\api{random} library is used to generate random numbers.
\api{NumPy} is a math library mainly used for array calculations.
\api{sklearn} is an open-source Python machine learning library 
that provides a large number of tools for data mining and analysis.
\api{linecache} is used to read arbitrary lines from a file.
\api{TensorFlow} is a powerful open-source software library developed by the Google Brain team for deep neural networks.

\begin{figure}[htpb]
    \centering
    \includegraphics[scale=0.4]{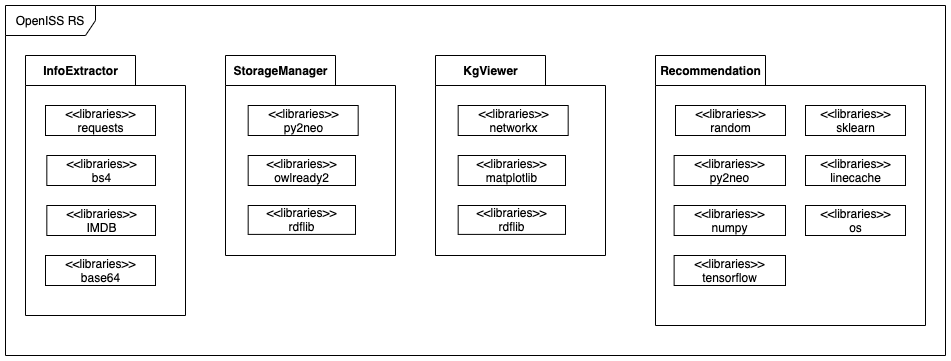}
    \caption{UML deployment diagram.}
    \label{fig:deployment_diagram}
\end{figure}

\section{Framework Application}

\subsection{Integrated Lenskit Application}
This Lenskit application is a comparison of NDCG (Normalized Discounted Cumulative Gain) values for Lenskit recommendation algorithms.
\xf{fig:lenskit-result} shows the result of the evaluation.

\begin{figure}[htpb]
    \centering
    \includegraphics[scale=0.4]{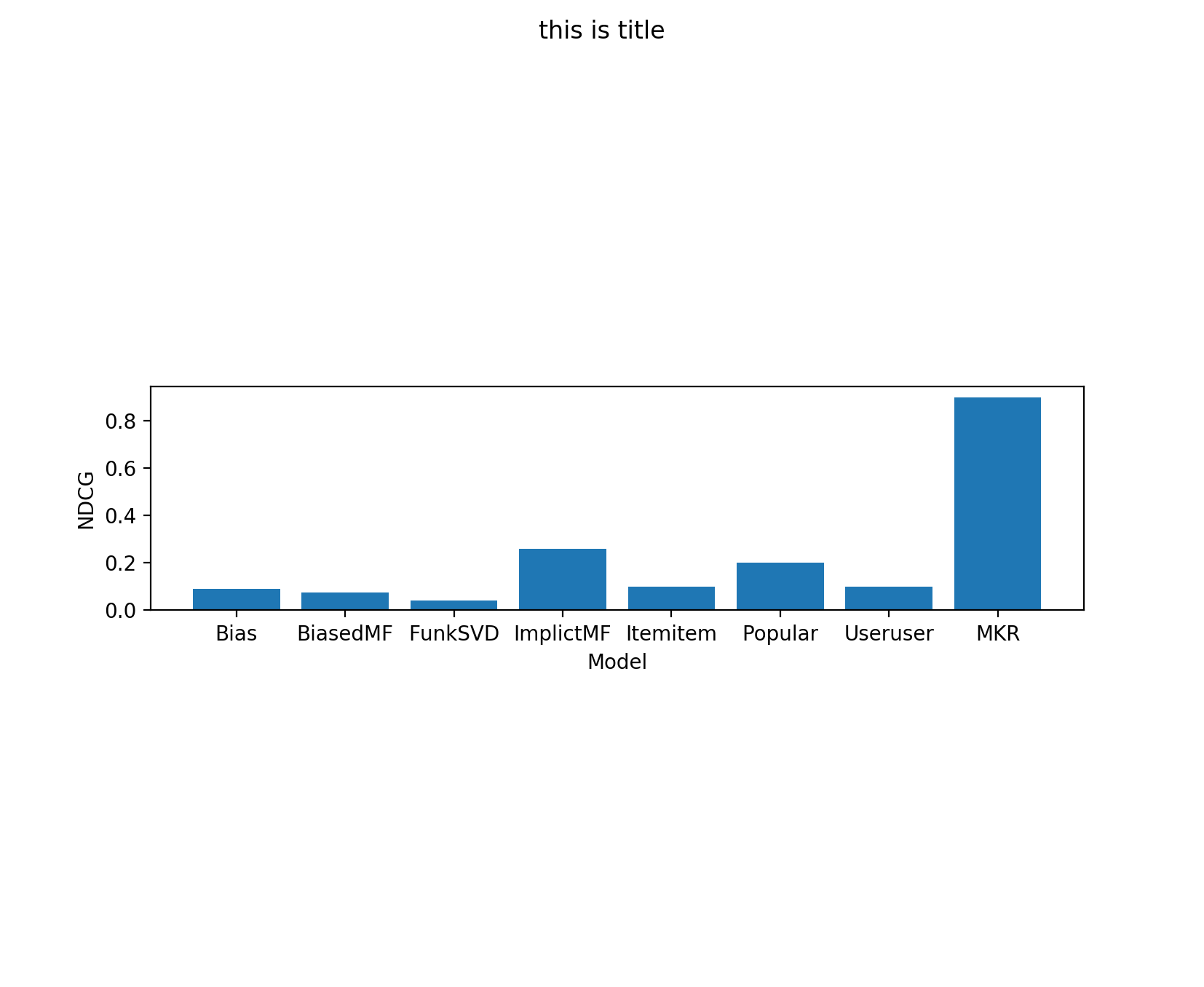}
    \caption {LensKit evaluation result.}
    \label{fig:lenskit-result}
\end{figure}

\subsection{Prediction of User's Rating for a Movie}
\label{sec:fw-app-reid}

So far, what we have described in the previous sectiuons are the design, implementation and instantiation of our framework. 
Our prediction service predicts users’ ratings of products (as an example product we have chosen movies).
The workflow can be described as follows:
\begin{enumerate}
    \item Read all the triple data through \api{dataLoader} to get the corresponding information.
    
    \item Use these triples and the user's rating of the movie as input, 
    and train through the RS module.
    
    \item Load the trained model, and predict the user's rating of the movie.
\end{enumerate}

For the prediction task, we use the pipeline shown in \xf{fig:fw-app-prediction-pipeline}.

\begin{figure}[htpb]
    \centering
    \includegraphics[scale=0.4]{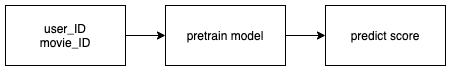}
    \caption[Prediction pipeline]
    {Prediction pipeline.}
    \label{fig:fw-app-prediction-pipeline}
\end{figure}

Now, with this pipeline instance, application developers no longer need to worry about details. 
All the user needs to do is provide the right input.

\subsection{Predict User's Rating for a Book}

The data we use is called \texttt{Book-Crossing} \cite{book-crossing}.
Book-Crossing dataset is Collected by Cai Nicolas Ziegler from the Book-Crossing community.

The Book-Crossing dataset comprises 3 tables. They are users, Books and Ratings.
Users contain the user's id, the user's location, and the user's age.
Books include book title, book author, year of publication, publisher and other information.
Ratings include user reviews of the book.
Because the process used is the same as \xs{sec:fw-app-reid}, we won't repeat it here.

\subsection{Knowledge Graph Fusion Application}

Knowledge fusion is an effective way to avoid node duplication in knowledge graphs. 
Users may have knowledge graph files in different formats which may result in duplicate nodes with the same information. We use knowledge graph fusion to solve the problem of duplicate nodes.

In our implementation, we provide an API where users can convert Neo4j triples to RDF 
or convert them to Neo4j based on RDF triples according to their needs. 

The procedure is described in Algorithm \ref{algo:fw-app-fusion}.

\begin{algorithm}
    parser $\leftarrow$ init a argument parser\;
    mode $\leftarrow$ init default storage mode\;
    path $\leftarrow$ init default path\;
    \;
    \If {args.mode == "neo4j2RDF"}
    	{call Neo4j to RDF API\;}
    \Else
	{call RDF to Neo4j API\;}
 
    \;
    Neo4j\_triples\_list $\leftarrow$ init a list for all the triples in Neo4j\;
    RDF\_triples\_list $\leftarrow$ init a list for all the triples in RDF\;
    \While{read file}{
    	add triples to Neo4j\_triples\_list \;
	add triples to RDF\_triples\_list \;
    	\While{read Neoj\_triples\_list or RDF\_triples\_list}{
		\If {node exist in RDF\_triples\_list or Neo4j\_triples\_list}{
		{find the corresponding node}
		
		\Else
		{create new node}
	}
		{connect nodes and relations}}
    }
    \caption{Knowledge graph fusion application procedure}
    \label{algo:fw-app-fusion}
\end{algorithm}

\section{Framework Evaluation and Conclusions}
\subsection{Evaluation Testbed Specifications}

Before starting the discussion about the evaluation, 
we first describe the environment, including the operating system used, processor power, memory, hardware, etc. 
The detailed specifications can be seen in \xt{tab:eval-laptop-server}.

\begin{table}[htpb]
    \centering
    \begin{tabular}{|c|c|c|c|}
        \hline
        \textbf{Setting} & \textbf{Name} & \textbf{Device}\\ \hline
        \multirow{3}{*}{Laptop} & Memory & 8 GB \\ \cline{2-3}
         & Processor & 2.3 GHz Intel Core i5\\ \cline{2-3}
         & Graphic & Intel Iris Plus Graphics 640 1536 MB \\ \cline{2-3}
         & OS & Mac OS Mojave 10.14.6\\ 
         \hline
        
       \multirow{3}{*}{Server (colab)} & Memory & 12 GB \\ \cline{2-3}
         & Processor & Intel Core i7-920 CPU 2.67GHz \\ \cline{2-3}
         & Graphic & GeForce GTX1080 Ti (12 GB) \\ \cline{2-3}
         & OS & Ubuntu 18.04.5 LTS 64-bit \\ 
         \hline
    \end{tabular}
    \caption{Environment hardware specifications.}
    \label{tab:eval-laptop-server}
\end{table}

\xt{tab:eval-libraries} lists the various libraries used in this research.

\begin{table}[htpb]
    \centering
    \begin{tabular}{|c|c|c|c|}
        \hline
        \textbf{Type} & \textbf{Name} & \textbf{Version}\\ \hline
        \multirow{9}{*}{Libraries} & py2neo & 4.3.0 \\ \cline{2-3}
        & \tool{Tensorflow} & 1.14.0\\ \cline{2-3}
        & \tool{bs4} & 4.8.0 \\ \cline{2-3}
        & \tool{csv} & 1.0 \\ \cline{2-3}
        & \tool{networkx} & 2.4 \\ \cline{2-3}
        & \tool{matplotlib} & 3.1.2 \\ \cline{2-3}
        & \tool{rdflib} & 4.2.2 \\ \cline{2-3}
        & \tool{numpy} & 1.17.0 \\ \cline{2-3}
        & \tool{sklearn} & 0.21.3 \\ \cline{2-3}
        & \tool{pandas} & 0.25.0\\ 
        \hline
    \end{tabular}
    \caption{Python libraries used.}
    \label{tab:eval-libraries}
\end{table}

Due to different operating systems, the software is slightly different, 
so we also need to give the software details of the environment. 
For these two environments: one is a laptop and the other is a server, 
we call them Setting~1 and Setting~2, respectively. 
There is a slight difference between the installed software versions. 
For these two environments, we give more detailed information in \xt{tab:eval-softwares}.

\begin{table}[htpb]
    \centering
    \begin{tabular}{|c|c|c|c|}
        \hline
        \textbf{Software} & \textbf{Setting 1} & \textbf{Setting 2}\\ \hline
        Python &  3.7.5 & 3.7.5  \\ \cline{2-3} \hline
        Neo4j Desktop & 1.2.1 & 1.2.9 \\ \cline{2-3} \hline
        Protege & 5.5.0 & 5.5.0 \\ \cline{2-3} \hline
    \end{tabular}
    \caption{Software packages and IDE tools used.}
    \label{tab:eval-softwares}
\end{table}

\subsection{Real-time Response}
According to \cite{realtime-rs}, we set the real-time response baseline to be 2000 ms.
To evaluate the whole system's processing ability, we performed the experiment described below:
\begin{enumerate}
    \item Load the trained model to get the pre-trained graph structure and weights. 
    \item Prepare feed\rule[-2pt]{0.2cm}{0.5pt}dict, new training data or test data. In this way, the same model can be used to train or test different data..
    \item Measure the difference between the timestamp $t_s$before the pipeline start and the timestamp $t_d$ after system processing.
    \item Repeat the previous operation 100 times, and then calculate the average processing time through the formula \autoref{eq:time_average}.
\end{enumerate}
\begin{equation}
\label{eq:time_average}
\mathit{Result} = \frac {\sum_{i = 1} ^{100} {t_e - t_s}}{100}.  
\end{equation}
The result shows that the speed of our solution on a local laptop machine is \texttt{634.33ms}. It is faster than the real-time baseline of 2000ms.

\subsection{Experimental Results}
We trained the MKR model using the MovieLens-1m dataset,
and then used the validation set to verify the model. 
We split all data according to 6:2:2, i.e., 60\% is the training set, 
20\% is the validation set, and 20\% is the test set. 
The data of the validation set and test set were not be used for training.

As \xs{sec:kg-storage-inst} shows, our side information has many types, 
including movie information, user information, and movie posters. 
We train with different side information through our model and 
obtain the results through 20 epoch training, as shown in \xt{tab:acc_auc_result}.
\begin{table}[]
    \centering
    \resizebox{\textwidth}{!}{
    \begin{tabular}{|c|c|c|c|c|c|c|}
        \hline
        \textbf{MovieLens 1M} & \textbf{train AUC} & \textbf{train ACC} & \textbf{evaluate AUC} & \textbf{evaluate ACC} 
        & \textbf{test AUC} & \textbf{test ACC}\\ 
        \hline
	{baseline} & 0.9189 & 0.8392 & 0.9046 & 0.8277 & 0.9042 & 0.8261 \\ \hline
	{baseline+movie} & 0.9227 & 0.8439 & 0.9081 & 0.8295 & 0.9061 & 0.8297 \\ \hline
	{baseline+user} & 0.9238 & 0.8455 & 0.9096 & 0.8321 & 0.9091 & 0.8331 \\ \hline
	{baseline+user+movie} & 0.9292 & 0.8516 & 0.9142 & 0.8375 & 0.9136 & 0.8359 \\ \hline
	{baseline+poster} & 0.9173 & 0.8279 & 0.9041 & 0.8153 & 0.9029 & 0.8214 \\ \hline
	{baseline+movie+user+poster} & 0.9273 & 0.8497 & 0.9113 & 0.8351 & 0.9111 & 0.8349 \\ \hline
    \end{tabular}}
    \caption{Results of the same models on different datasets.}
    \label{tab:acc_auc_result}
\end{table}

From the results, we can see that the accuracy of data with user information and movie information is the highest, about 1\% higher than the baseline. 
Because users may watch a movie because of a director or an actor,
the other movie information can help improve accuracy.
The age, job and other information in the user information also help to improve the accuracy, 
because the user may choose some related movies to watch based on age and occupation.
But because the poster of each movie is different, 
in the knowledge graph, each poster is connected to only one movie node, 
so the poster data is sparse data for the knowledge graph. 
Therefore, the poster information does not have a good effect for us at present, 
but if we can extract some useful information from the poster through technologies such as computer vision, 
it may be helpful in improving the accuracy of the recommendation.

\subsection{Concluding Remarks and Extensions}
We proposed, designed and implemented a generic software framework 
that integrates knowledge graph representations for providing training data to recommendation methods. 
At the core of this framework are knowledge graphs for storing and managing information of use to recommendation methods. To the best of our knowledge, a similar framework is not available elsewhere. 

The ultimate goal of our work is to make it as a research platform for more developers in the recommendation systems field. With that goal it needs to be extended as follows:

\textbf{Java API wrapper:} 
Our framework was written in Python, 
but the movie recommendation system is mostly used on web pages.
So it is better for users, if can we provide a Java wrapper for our API.

 \textbf{Support different machine learning backend:}
Currently, our recommended module only supports TensorFlow. But there are many different deep learning frameworks, such as PyTorch, Caffe or Scikit-learn. Different frameworks have their advantages. We plan to add various machine learning frameworks to our framework in the future.

 \textbf{Support more storage methods and more input formats:} 
Currently, we only support four storage formats, namely RDF, RDFS, OWL and Neo4j. 
For the input format, because we use CSV for storage, some users may choose JSON format or TTL format, so we also need to update the program to support these formats.



\bibliographystyle{elsarticle-num} 
\bibliography{openiss-rs-kg-framework}

%
%
%
%
\end{document}

%% file: commands.tex
\newcommand{\xf}[1]{Figure~\ref{#1}}

\newcommand{\xs}[1]{Section~\ref{#1}}

\newcommand{\xt}[1]{Table~\ref{#1}}

\newcommand{\file}[1]{\url{#1}\index{Files!#1}}
\newcommand{\tool}[1]{\texttt{#1}\index{Tools!#1}}

\newcommand{\api}[1]{\texttt{#1}\index{API!#1}}

\newcommand{\lucidL}[1]{{$\mathit{Lucid}$}($L$) }

		{}

\def\myvert{\raise 2.27pt \hbox{\vrule depth 0pt height 8pt width 0.2mm}}
\def\myarrow{\hspace*{0.43mm}%
             \raise 2.29pt\hbox{\vrule depth 0pt height 8pt width 0.16mm}%
             \hspace*{-0.32mm}%
             $\longrightarrow$
             \ %
             }
